\documentclass[aps,prl,twocolumn,groupedaddress,showpacs]{revtex4}

\usepackage{amsmath}
\usepackage{microtype}
\usepackage{subfigure}
\usepackage{amsfonts}
\usepackage{amssymb}
\usepackage{changepage}
\usepackage{graphicx}

\newcommand{\be}{\begin{equation}}
\newcommand{\ee}{\end{equation}}

\newcommand{\rd}{{\rm d}}

\newcommand{\lp}{\ell_{\rm p}}

\newcommand{\li}{\ell_0}

\newcommand{\kbt}{k_{\rm B}T}

\newcommand{\ve}{\varepsilon}

\newcommand{\sD}{{\sf D}}

\newcommand{\sK}{{\sf K}}

\newcommand{\lc}{{\ell_\mathrm{c}}}

\newcommand{\Ue}{U_{\rm ext}}

\newcommand{\dUe}{\dot U_{\rm ext}}
\newcommand{\kt}{\kappa^{(2)}}
\newcommand{\kd}{\kappa^{(3)}}
\newcommand{\ktr}{\kappa^{(t)}}
\newcommand{\kmd}{\kappa^{({\rm MD})}}
\newcommand{\ddx}[2]{\frac{\partial #1}{\partial #2}}

\newcommand{\dx}[1]{\mathrm{d} #1}
\newcommand{\DDx}[2]{\frac{\dx{#1}}{\dx{#2}}}
\newcommand{\DDxn}[3]{\frac{\mathrm{d}^#3 #1}{\mathrm{d} #2^#3}}

\def\lsim{\mathrel{\rlap{\lower4pt\hbox{$\sim$}}
    \raise1pt\hbox{$<$}}}                % less than or approx. symbol
\def\gsim{\mathrel{\rlap{\lower4pt\hbox{$\sim$}}
    \raise1pt\hbox{$>$}}}           

\begin{document}
\preprint{APS/123-QED}

\title{Catch Bonding in the Forced Dissociation of a Polymer Endpoint}

\author{Cyril Vrusch}
%\affiliation{
% Department of Applied Physics and Institute for Complex Molecular Systems, Eindhoven University of Technology,\\
% P.O. Box 513, NL-5600 MB Eindhoven, The Netherlands
%}%
\author{Cornelis Storm}
\affiliation{
 Department of Applied Physics and Institute for Complex Molecular Systems, Eindhoven University of Technology,\\
 P.O. Box 513, NL-5600 MB Eindhoven, The Netherlands
}%


%\affiliation{
%\affiliation{
%\affiliation{%


\date{\today}% It is always \today, today,

\begin{abstract}
Applying a force to certain supramolecular bonds may initially stabilize them, manifested by a lower dissociation rate. We show that this behavior, known as catch bonding and by now broadly reported in numerous biophysics bonds, is generically expected when either or both the trapping potential and the force applied to the bond possess some degree of nonlinearity. We enumerate possible scenarios, and for each identify the possibility and, if applicable, the criterion for catch bonding to occur. The effect is robustly predicted by Kramers theory, Mean First Passage Time theory, and finally confirmed in direct MD simulation. Among the catch scenarios, one plays out essentially any time the force on the bond originates in a polymeric object, implying that some degree of catch bond behavior is to be expected in {\em any} protein-protein bond, as well as in more general settings relevant to polymer network mechanics or optical tweezer experiments.

%\begin{description}.
%\end{description}
\end{abstract}

\maketitle
How long does it take a fluctuating particle to escape the trap of a confining potential well? The question is one of the staples of statistical mechanics and, in its simplest incarnation, gives rise to Kramers' well-known expressions for the rate at which a particle crosses a potential energy barrier---the rate exponentially decaying with increasing barrier height \cite{Kramers1940}. This {\em escape problem} features in a wide range of problems in statistical mechanics, and has important applications and consequences in materials science, soft matter- and biological physics for its capacity to predict, under general external conditions, the dissociation rate of bound states and, thereby, the mean lifetime $\langle \tau \rangle$ of bonds between, for instance, receptor-ligand pairs \cite{Merkel1999,Escude2014,marshall2003direct}.

Escape kinetics change when forces are taken into consideration. Generally, an applied force aligning with the escape path will hasten dissociation; the naive Kramers prediction is, again, an exponential decrease in lifetime with increasing force \cite{Kramers1940}. As \cite{Suzuki2011} showed, forced escape scenarios become considerably richer when the energy landscape is multidimensional---in particular, they note the curious possibility of the lifetime of the bound state initially {\em increasing} with the applied force, a phenomenon they termed {\em rollover}. This counter-intuitive behavior---a bond is strengthened by applying a force to it---is no longer a theoretical curiosity but has, in recent years, been experimentally demonstrated in a range of non-covalent biophysical bonds where it has become widely known as a {\em catch bond} \cite{Marshall2003,Kong2009}. For protein-protein bonds, such behavior is generally ascribed to specific conformational properties of the molecules involved \cite{VogelRev,WThomasReview,lou2007structure,sarangapani2011regulation,chen2011polymer}; we will show that it is generic. In this paper, we will define a catch bond to mean a bound state whose average lifetime $\langle \tau \rangle(f)$ possesses an initial regime of increase with force:

\be
\left.\frac{{\rm d}}{{\rm d} f}\langle \tau \rangle(f)\right|_{f=0}>0\, .
\ee 

\begin{figure}[t!]
\begin{center}
\includegraphics[width=\linewidth]{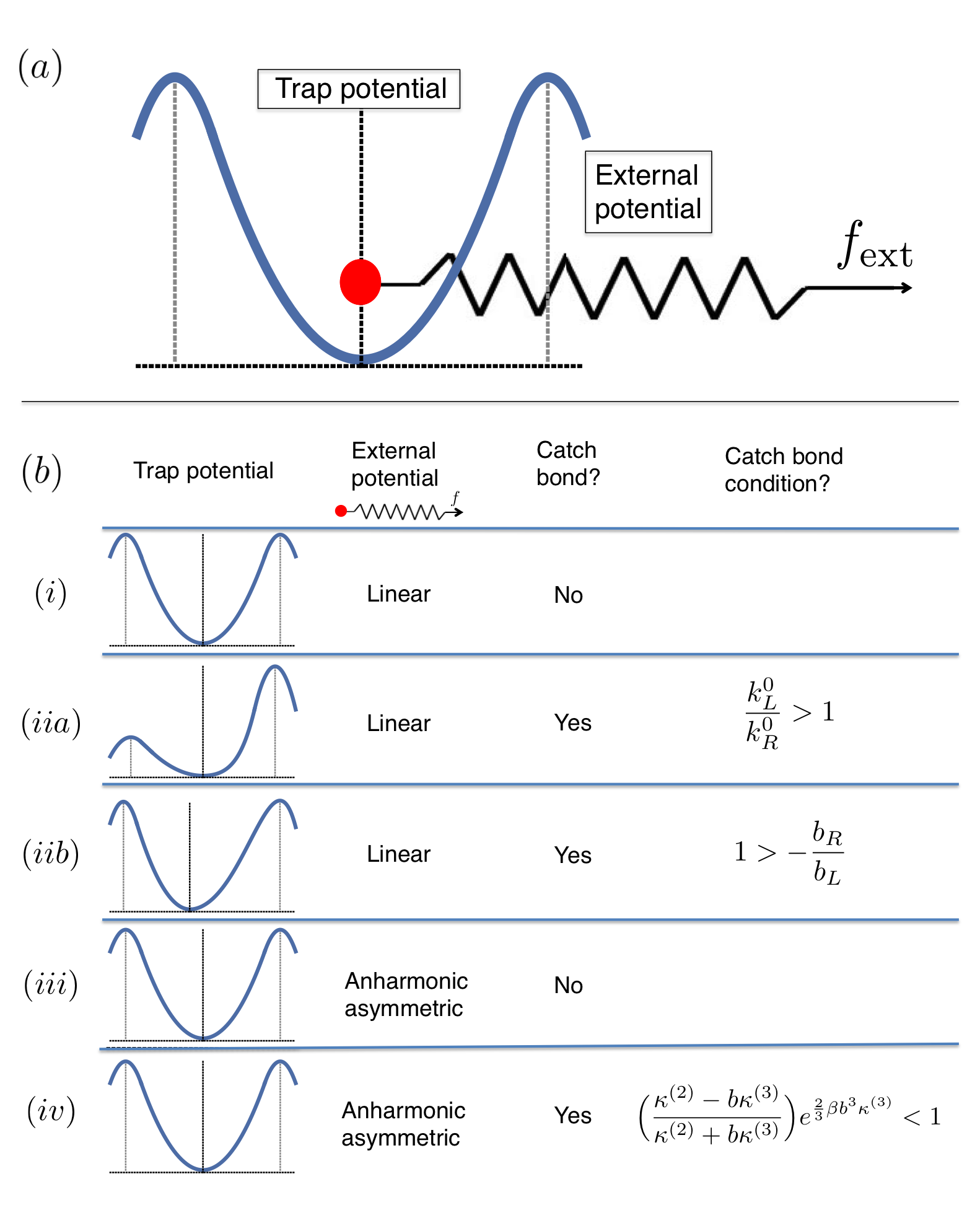}
\caption{\label{Fig1} (a) Illustration of the bond, the bead (red) is trapped in an potential well while an external potential acts on it. (b) Scenarios for the energy landscapes with corresponding catch bond criteria.}
\end{center} 
\end{figure}

What we will show, is that this behavior is generic when the applied force is non-linear (i.e., depends asymmetrically on the position of the particle in the trap), even in one-dimensional energy landscapes, and moreover even in symmetric one-dimensional potentials. Our findings demonstrate the broad generality, without the need for further assumptions, of the implication of entropic elasticity in catch bonding \cite{wei2008entropic}. We also show, that this scenario plays out essentially any time the force is transmitted to the escaping particle by a polymeric object, and thus that some regime of catch bond behavior is to be expected in {\em any} protein-protein bond, as well as more general settings relevant to polymer network mechanics or optical tweezer experiments. 

Our paper is organized as follows: First, we outline the general framework of a forced one-dimensional escape process in the Kramers sense (with rates depending only on barrier height) and summarize the various trapping and forcing scenarios that give rise to catch bonding. Next, we specify to the case of a particle that is pulled out of a confining trap by a polymer under tension, and study the full differential equation determining the mean first passage time (MFPT) for a general asymmetric potential. Our explicit analytical solution proves, that a nonlinear force-extension relation for the polymer is a necessary condition for catch bonding to occur. We conclude our paper with a Molecular Dynamics (MD) simulation of the escape of a tensed semiflexible polymer with one end trapped in a symmetric, harmonic potential well showing significant (approximately 10\% increase in unbinding time) catch bonding, while also highlighting some of the more subtle features of polymer unbinding.
  
{\it Kramers Catch Bonding Scenarios}
An intuitive model to rationalize equilibrium catch bond behavior is the so-called Two-Pathway Model (TPM), developed by Pereverzev {\it et al.}\cite{Pereverzev2005}. A particle is confined to a local minimum of the trap's potential energy $U_t(x)$, where $x$ is the position of the particle in the trap. This local minimum $U_t=0$ (the bound state) is flanked by two different energy barriers. In any trap configuration, there are two escape paths; one left ($L$) and one right ($R$). By applying a force $f$ that is orthogonal to the pathway with the lowest barrier, however, the effective barrier height in the secondary direction is lowered and escape along the second unbinding pathway becomes increasingly likely and frequent. Provided the dissociation lengths of the two pathways are suitably chosen, the same applied force may simultaneously increase the barrier height of the original pathway, making it less likely. The combined escape problem over the two barriers determines the overall escape time, and the differential response to the applied force may result in an increase in the average unbinding time as a force is applied, that is, catch bond behavior. As such, catch bonding in the TPM is a result of an asymmetric trap, in combination with a linear forcing.

We generalize the one-dimensional TPM by making no assumptions on the energy barriers, but only assuming that there are two pathways by which the bond can unbind. We will continue to call those pathways left ($L$) and right ($R$), and shall call the dissociation lengths to the left and right $b_L$ and $b_R$, respectively. The bare escape rates corresponding to the pathways are called $k_L^0$ and $k_R^0$, and in the Kramers picture are determined by the barrier heights presented by the trap potential only
\be
k_i^0=\nu\, e^{-\beta U_t(b_i)}, \quad (i=L,R),
\ee
with $\nu$ the attempt frequency and $\beta=(\kbt)^{-1}$. For two otherwise independent processes, we thus expect the combined rate of escape to be $k^0=k_L^0+k_R^0$. These rates define the survival probabilities $S_L(t)$ and $S_R(t)$---the probabilities that unbinding did {\em not} occur along the $L$ or $R$ pathway prior to time $t$---as
\be
S_L(t)=e^{-k_L^0 t}\, ; \, S_R(t)=e^{-k_R^0 t}\, ,
\ee
and permit to compute $\pi_L(t)\, \rd t$ ($\pi_R(t)\, \rd t$), the likelihood that the {\em first} unbinding event occurs along $L$ ($R$) between $t$ and $t+\rd t$, as the joint probability of unbinding along $L$ ($R$) between those times, and having survived unbinding along $R$ ($L$) up until $t$:
\be
\pi_L(t)=-\DDx{S_L(t)}{t}\cdot S_R(t)\, ; \, \pi_R(t)=-\DDx{S_R(t)}{t}\cdot S_L(t)\,.
\ee
The asymptotic probabilities $P_L$ and $P_R$ of unbinding along either direction are then given by
\begin{eqnarray}
P_L&=&\int_0^\infty \pi_L(t)\, {\rd t}\,=\frac{k_L^0}{k_L^0+k_R^0}\, ,\\
P_R&=&\int_0^\infty \pi_R(t)\, {\rd t}\,=\frac{k_R^0}{k_L^0+k_R^0}\, .
\end{eqnarray}
This makes intuitive sense; if one of the two rates is very high compared to the other, fast unbinding preempts dissociation along the slow pathway. It is, however, also instructive to consider the average unbinding times, separately for each of the two directions. The mean unbinding times along $L$ and $R$ are computed as
\begin{eqnarray}
\langle \tau_L^0 \rangle &=& \int_0^\infty t \left(\frac{\pi_L(t)}{P_L}\right)\, {\rd t}\,=\frac{1}{k_L^0+k_R^0}\, ,\\
\langle \tau_R^0 \rangle &=& \int_0^\infty t \left(\frac{\pi_R(t)}{P_R}\right)\, {\rd t}\,=\frac{1}{k_L^0+k_R^0}\, ,
\end{eqnarray}
where the factors $P_L$ and $P_R$ are included to normalize the $L$ and $R$ unbinding distributions; $\langle \tau_L \rangle$ and $\langle \tau_R \rangle$ are thus the mean unbinding times of each path, averaged over only the unbinding events in that particular direction. This reveals two interesting properties: the presence of a second unbinding path always decreases the average unbinding lifetime along a primary path (the unbinding is, of course, as least as fast as it was originally) but the average time it takes to escape along either direction becomes the same. Counter-intuitively, perhaps, the escape along a very slow direction is greatly sped up by the presence of a secondary, fast unbinding route; this is because only the rare, very quick event along the slow path is not precluded by the abundant, fast rebinding. The overall number of such events may become very small, though, as borne out by $P_L$ and $P_R$. Averaging the escape time over all events, both $L$ and $R$, gives the expected result mean lifetime $\langle \tau \rangle$ of the bond
\be\label{tauav}
\langle \tau^0 \rangle=P_L\langle \tau_L^0 \rangle+P_R\langle \tau_R^0 \rangle=\frac{1}{k_L^0+k_R^0}= \frac{1}{k^0}\, .
\ee
Thus, we find that due to the presence of the other pathway, each path acquires the same escape time. This escape time, moreover, is equal to the overall lifetime of the bound state 
\be
\langle\tau^0\rangle=\langle\tau_L^0\rangle=\langle\tau_R^0\rangle\, .
\ee 

How do external manipulations, such as the application of a force, affect $\langle\tau\rangle$? Let us consider now the case in which the unbinding process takes place in the presence of an additional, external potential $\Ue(x,\ve)$ which defines a (possibly position-dependent) force $f_{\rm ext}(x,\ve)=-{\rd \Ue}/{\rd x}$. The parameter $\ve$ quantifies some manner of tuning this external potential; in the simplest case it is the force itself ($\Ue(x,\ve)=-\ve x$), but it may also represent some more general way of altering the externally applied potential. Assuming that the rate at which the particle crosses the boundaries $b_R$ or $b_L$ of the trap still depends on the barrier height only (we will examine the validity of this assumption in the latter part of this article), $\Ue(x,\ve)$ shifts the transition rates according to
\be\label{keq}
k_i=k^0_i \, e^{-\beta\, \Ue(b_i,\ve)}, \quad (i=L,R).
\ee
To assess the possibility of catch bonding, we compute the change in the overall lifetime $\langle \tau \rangle$
\be
\langle \tau \rangle(\ve)=\langle \tau \rangle(\ve=0)+\langle \Delta \tau \rangle(\ve)\, ,
\ee
due to a small change in $\ve$. This is done by substituting Eq. (\ref{keq}) into Eq. (\ref{tauav}) and expanding to lowest (first) order in $\ve$, to yield

\be
\begin{aligned}
\langle \Delta \tau \rangle(\ve)&= \beta \ve \left(\frac{e^{\beta(\Ue(b_L,0)+\Ue(b_R,0))}}{(e^{\beta\Ue(b_R,0)}k_L^0+e^{\beta\Ue(b_L,0)}k_R^0)^2}\right)\times \\
&\times\left( e^{\beta\Ue(b_R,0)}k_L^0 \left[\left.\ddx{\Ue}{\ve}\right|_{(b_L,0)}\right]+\right. \\
&+\left.e^{\beta\Ue(b_L,0)}k_R^0 \left[\left.\ddx{\Ue}{\ve}\right|_{(b_R,0)}\right] \right)+{\cal O}(\ve^2)\, .
\end{aligned}
\ee
The essential information in this cumbersome expression may be condensed somewhat. The first factor, between brackets, is strictly positive and does not affect the sign of $\langle \Delta \tau \rangle(\ve)$. For $\langle \Delta \tau \rangle$ to increase with increasing $\ve$, the following 'catch criterion' must hold
\be\label{crit}
\frac{k^0_L}{k^0_R}>-\left(\frac{e^{\beta\Ue(b_L,0)}}{e^{\beta\Ue(b_R,0)}}\right)\left(\frac{\dUe(b_R,0)}{\dUe(b_L,0)}\right)\, .
\ee 
Note, first, that the left hand side is entirely set by the trap potential $U_t(x)$, while the right hand side is determined only by $\Ue(x)$. $\dUe$ is shorthand for the $\ve$-derivative of $\Ue$, which is to be taken before $\ve$ is set to zero. Either $U_t$ or $\Ue$, or indeed both, may be used to create a catch bond effect. This equation thus encodes a number of scenarios as illustrated in Fig.\ref{Fig1}:

{\em (i): Symmetric trap, constant external force}. $U_t(x)$ is symmetric (that is, $U_t(x)=U_t(-x)$), and $\Ue(x,\ve)=-\ve x$ (that is, the external force $f=\ve$ is constant for given $\ve$, and points in the $R$-direction throughout the trap region $[b_L,b_R ]$). In this case, because of the symmetry of $U_t$ it must be true that $b_L=-b_R$, as well as $k^0_L=k^0_R$. Since $\Ue(x,0)=0$, and the $\ve$-derivatives at the boundaries are equal but opposite, the catch criterion cannot be met; both sides of Eq. (\ref{crit}) are 1.

{\em (ii): Asymmetric trap, constant external force}. $U_t(x) \neq U_t(-x)$, and $\Ue(x)=-\ve x$ in $[b_L,b_R ]$. An asymmetry in $U_t$ may be relevant in two different ways; either ({\em iia}) the dissociation lengths $b_L$ and $b_R$ to either side are equal in absolute magnitude, but $U_t(b_L)\neq U_t(b_R)$, that is, the barrier heights are different; or ({\em iib}) the dissociation lengths differ ($b_L \neq b_R$) while the trap barrier heights are the same $U_t(b_L)= U_t(b_R)$. In case ({\em iia}), calling the dissociation length $-b_L=b_R=b$, the criterion for increased lifetime reduces to
\be
\text{(\em iia)}:\quad\frac{k_L^0}{k_R^0}>1\, .
\ee
This is the TPM catch bonding scenario: if an external force opposing escape along the initially favored (most frequent) pathway is applied, the overall escape rate increases. However, also in case ({\em iib}) catch bonding may occur; in this case, the criterion reduces to
\be
\text{(\em iib)}:\quad1>-\frac{b_R}{b_L}\, .
\ee
Thus, even for identical barrier heights applying a pulling force in the direction where the dissociation length is shortest will increase the overall lifetime of the bond. Obviously, suitably chosen combinations of trap asymmetries ({\em iia}) and ({\em iib}) also produce catch bonding.

{\em (iii): Symmetric trap, harmonic external potential}. For symmetric $U_t(x)$, and thus again $-b_L=b_R=b$ and $k^0_L=k^0_R$, the left hand side of Eq. (\ref{crit}) is equal to one. Suppose now, that the external potential is supplied by a harmonic spring of rest length $\li$ and spring constant $\kt$, whose left end point is attached to the particle in the trap, and whose right end is at $x=\li(1+\ve)$. $\ve$ thus measures the extension of the spring and relates to the tension in the spring. In that case, the external potential (with $x=0$ the center of the trap) is given by
\be\label{harm}
\Ue(x,\ve)=\frac{\kt}{2} (\ve\li-x)^2\,.
\ee
Straightforward substitution into Eq. (\ref{crit}) yields that in this case, too, the catch criterion cannot be met, as both the left- and right hand sides of Eq. (\ref{crit}) are 1. Moreover, expanding the change in lifetime to second order in $\ve$ reveals that $\langle \Delta \tau \rangle(\ve)\sim - \ve^2$, and thus suggests that the lifetime at zero displacement (i.e., zero force applied to the spring) is a maximum.

{\em (iv): Symmetric trap, anharmonic external potential}. We may also consider the same trap conditions as case ({\em iii}), but with an anharmonic, nonlinear external potential (we will use a third power here, but the argument is valid with the inclusion of arbitrary odd powers) 
\be\label{anharm}
\Ue(x,\ve)=\frac{\kt}{2} (\ve\li-x)^2+\frac{\kd}{3} (\ve\li-x)^3\,.
\ee
In this case, the catch criterion may be cast as
\be
\left(\frac{\kt-b\kd}{\kt+b\kd}\right)e^{\frac23 \beta b^3 \kd}<1\, ,
\ee
demonstrating that for suitably chosen nonlinearity $\kd$ catch bonding is possible. In particular, one recognizes the limits $\kd \to 0$ (recovering case ({\em iii})) to preclude catch bonding, whereas $\kt \to 0$ will always give catch bonding, for all positive values of $\kd$.

Summarizing, a range of configurations of external forcing and trap properties give rise to an increase in lifetime with applied force (or with the extension of a springlike tether to which the escaping particle is attached). In particular, we show here that such a regime is generic for nonlinearly elastic tethers. Examples of such are all polymers, flexible and semiflexible. Most proteins also display highly nonlinear force-extension relationships, and our result suggests a route towards catch binding that is not due to any specific allosteric or conformational mechanism, but rather is encoded within the universal departures from nonlinearity in protein mechanics. To explore this mechanism in more detail, we focus on cases {\em (iii)} and {\em (iv)} in the following.

The Kramers analysis assumes $L$ and $R$ transition rates to depend only on barrier height, and ignores the shape of the potential along the different unbinding pathways. In order to account for those, and check the robustness of the Kramers predictions, we now compute the lifetimes via the full mean first passage time formalism.

{\em Mean first passage time}
The average time for the particle to escape the trap, and thus for dissociation of the bond, is the mean first passage time (MFPT) for the particle to pass the boundaries of the trap. The MFPT $\langle\tau(x)\rangle$, with $x$ the point of departure inside the trap, obeys the differential equation \cite{Redner2001}   
\be 
\frac{f_{\rm tot}(x)}{\kbt}\frac{\mathrm{d}}{\mathrm{d}x}\langle\tau(x)\rangle+\frac{\mathrm{d}^2}{\mathrm{d}x^2}\langle\tau(x)\rangle=-\frac{1}{\sD}\, ,\label{MFPT}
\ee
\\
supplemented with the boundary condition that $\langle\tau(x)\rangle=0$ at the boundaries $b_L$ and $b_R$ of the trap. Here is $\sD$ the diffusion constant for the fluctuating particle, and $f_{\rm tot}(x)$ is the total force acting on the bead, defined as $-\DDx{U_{\rm tot}}{x}$ with $U_{\rm tot}(x)=U_t(x)+\Ue(s)$.

We will focus on potential catch bonding for symmetric traps, in combination with harmonic or anharmonic external forcing to mimick the situation where the escaping particle is actually one end of a polymeric object or protein. Thus, we choose as our trapping potential
\be   U_t(x)=\left\{
\begin{array}{ll}

      \frac{1}{2} \ktr(x^2-b^2) & |x| \leq b \\
      0 & x>b \\
\end{array}, 
\label{ExCrosslink}
\right.\ee
i.e., harmonic with trap spring constant $\ktr$ in the region $[-b,b]$, and zero elsewhere (this choice, as well as the discontinuity at the boundaries, is immaterial; the potential outside of the trap region is irrelevant to the escape problem). Escape is defined as first passage across the left or right boundary.

\begin{figure}[t!]
\begin{center}
\includegraphics[width=\linewidth]{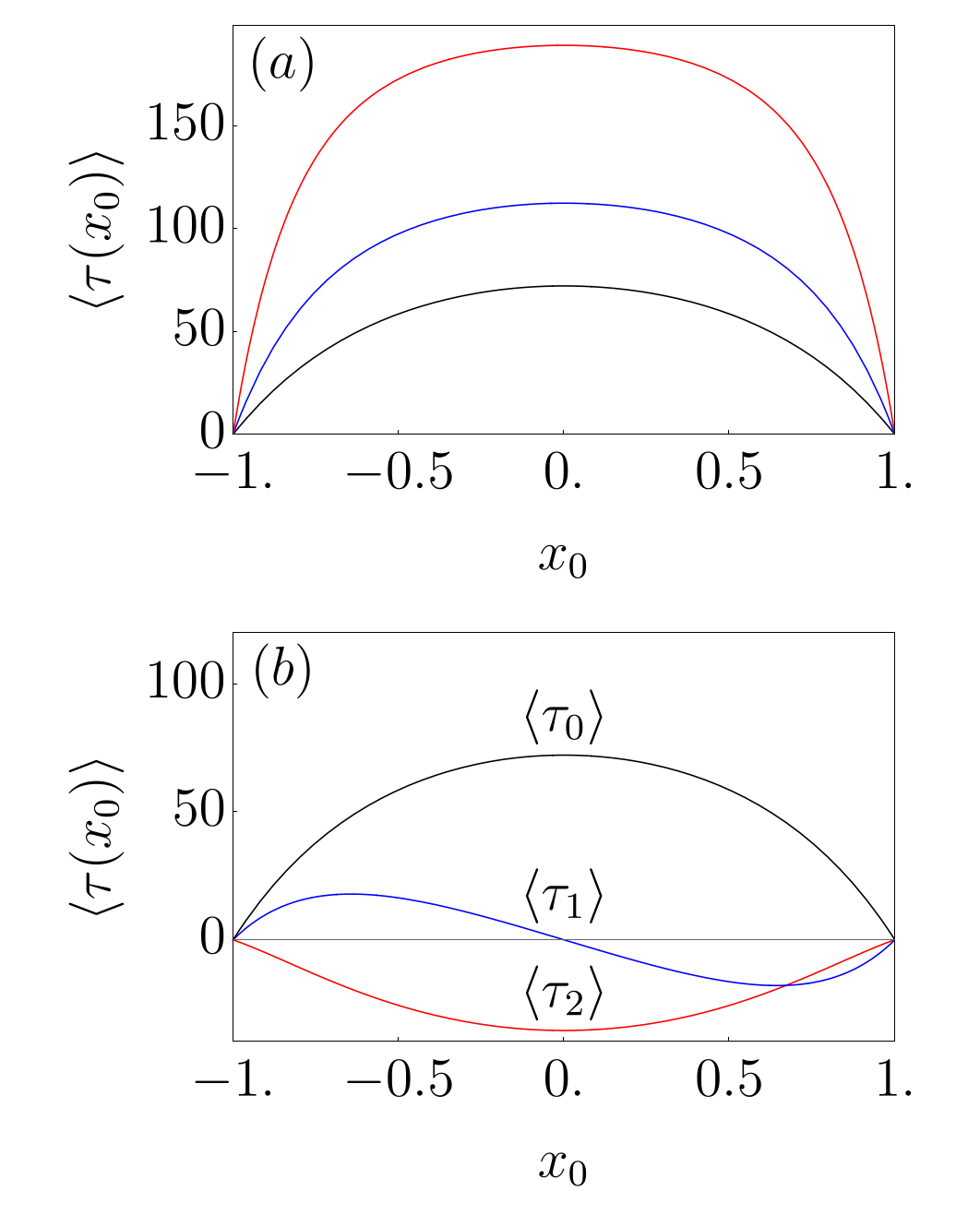}
\caption{\label{TauX} (a) The average unbinding time $\langle\tau(x_0)\rangle$ obtained by Eq. \ref{tau} for a crosslink which released at $x_0$ at $t=0$. The sum of $\kappa_c$ and $\kappa_s$ is 2 (black line), 4 (blue line), and 6 (red line). (b) The first three terms of terms of the perturbed system for $\kappa=2$.}
\end{center} 
\end{figure}

{\em (iii): Symmetric trap, harmonic external potential}
When the particle is attached to a harmonic spring with an energy of the of the form of Eq. (\ref{harm}), the total force on the particle is given by
\be\label{ftotharm}
f_{\rm tot}(x,\ve)=-\ktr x +\kt (\ve \li-x)
\ee
When we set $\ve=0$, the end point of the spring is fixed a distance $\ell_0$ away from the center of the trap and the total force becomes simply $f_{\rm tot}(x,0)=-(\ktr+\kt) x$. In this case, Eq. (\ref{MFPT}), with boundary conditions $\langle\tau_0(\pm b)\rangle=0$ may be solved analytically:
\be
\begin{aligned}
\langle\tau_0(x_0)\rangle & =\frac{b^2 \,_2F_2(\{1,1\},\{\frac{3}{2},2\},\frac{b^2(\ktr+\kt)}{2\kbt})}{2\sD}\\ & -\frac{x_0^2 \,_2F_2(\{1,1\},\{\frac{3}{2},2\},\frac{x_0^2(\ktr+\kt)}{2\kbt})}{2\sD},
\label{tau}
\end{aligned}
\ee
where $_pF_q(a;b;z)$ is the generalized hypergeometric function. $x_0$ is the position of the bead at $t=0$. $\langle\tau(x_0)\rangle$ is plotted in Fig. \ref{TauX}(a) for various values of $\kappa=\ktr+\kt$. As expected, the escape time is maximal when the particle departs from the center of the trap; we shall take this value $\langle\tau(0)\rangle$ at $\ve=0$ as our reference time.

To study the effect of putting the spring under tension, we now increase $\ve$ away from zero. This results in an extended tether in the center of the trap, and $f_{\rm tot}$ is now given by Eq. (\ref{ftotharm}). The effect of this small change may be studied perturbatively. When we expand the perturbed solution to second order in $\ve$ as
\be
\langle\tau_{\epsilon}(x_0)\rangle=\langle\tau_0(x_0)\rangle+\epsilon\,\langle\tau_1(x_0)\rangle+\epsilon^2\,\langle\tau_2(x_0)\rangle+{\cal O}(\ve^3),
\ee
with $\langle\tau_0(x_0)\rangle$ is the exact solution for $\epsilon=0$, $\langle\tau_1(x_0)\rangle$ and $\langle\tau_2(x_0)\rangle$ (and, indeed, the higher order corrections) may be obtained from an order-by-order set of recursive differential equations
\be
\begin{aligned}
&-\!\!\frac{x_0(\ktr+\kt)}{\kbt} \DDx{\langle\tau_n\rangle}{x_0}+\frac{\kt \li}{\kbt}\DDx{\langle\tau_{n-1}\rangle}{x_0}+\DDxn{\langle\tau_n\rangle}{x_0}{2}=0\, ,
\end{aligned}
\ee
with boundary conditions $\langle\tau_n(\pm b)\rangle=0$. These equations, too, may  be analytically solved for $\langle\tau_1(x_0)\rangle$ and $\langle\tau_2(x_0)\rangle$. Fig. \ref{TauX}(b) graphs these first two corrections to the $\ve=0$ result, and confirms what was already suggested by the Kramers analysis: $\langle\tau_1(0)\rangle=0$, meaning that there is no effect, to first order in $\ve$, on the lifetime of the bond (i.e., no catch bond). Moreover, we also see that indeed the second order correction $\langle\tau_2(0)\rangle=0$ is negative, proving that the zero-$\ve$ lifetime is indeed a maximum. Attaching the escaping particle to a linear spring cannot increase the lifetime in a symmetric potential.

\begin{figure}[b!]
\begin{center}
\includegraphics[width=\linewidth]{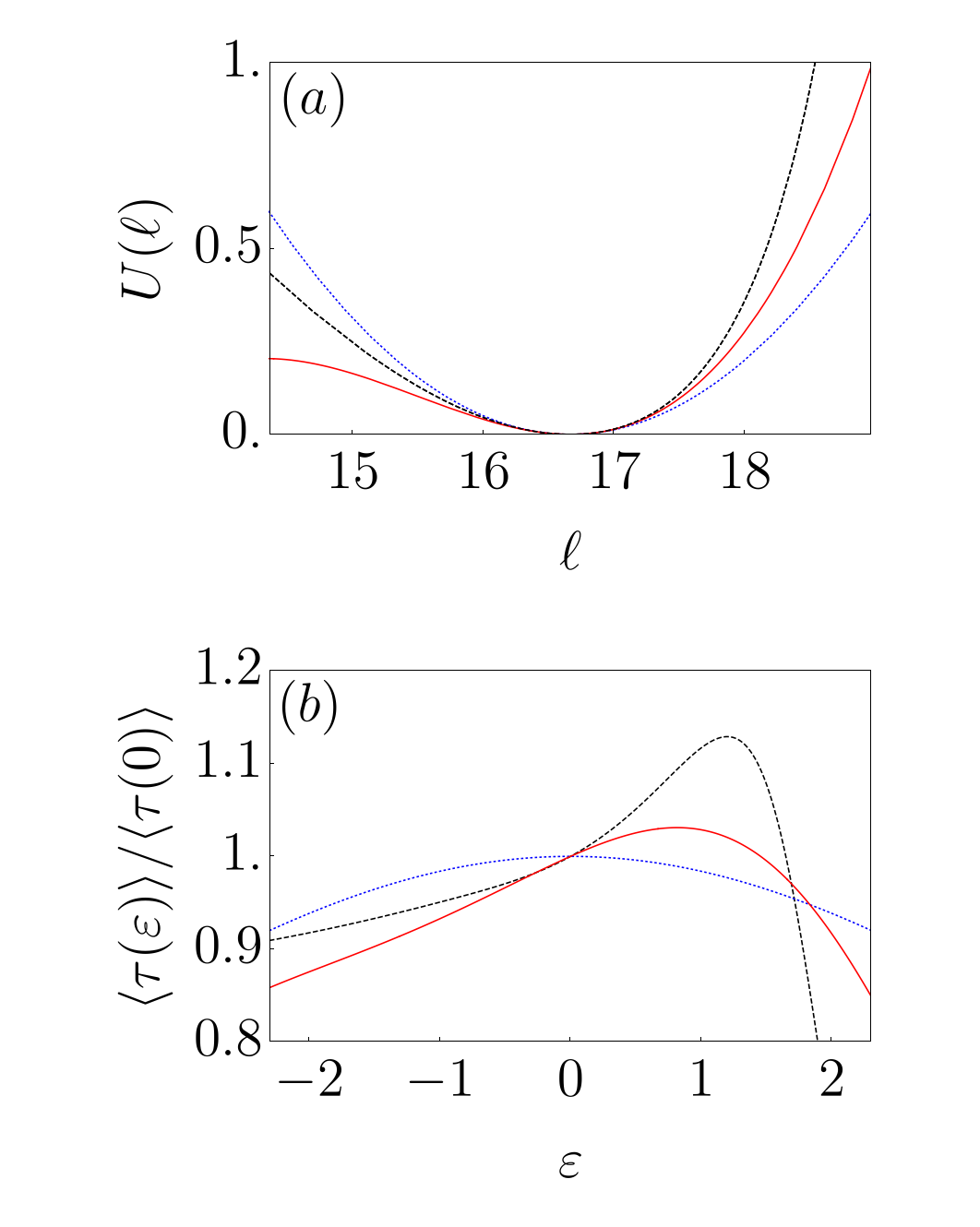}
\caption{\label{Fig3} (a) Potentials $\Ue(x,0)$ and (b) average unbinding times relative to their $\ve=0$ values for a particle attached to a semi-flexible fiber with $\ell_p = \ell_c=20$ (dashed black line), the first harmonic approximation to the WLC (dotted blue line) and the first anharmonic approximation (red line).}
\end{center} 
\end{figure}

{\em (iv): Symmetric trap, anharmonic external potential}
Kramers theory predicts catch bonding in the case of an anharmonic external potential. To connect this to a more realistic setting, we will specify to the case of a Worm-Like Chain tether; instead of a Hookean spring the escaping particle is now connected to a semiflexible polymer with a contour length $\lc$ and a persistence length $\lp$. The force-extension relation for such a polymer is strongly nonlinear, and given by \cite{Dennison2016,Huisman2008}
\be
\ell(f)=\lc-\frac{\kbt}{2f}\left[\sqrt{\frac{f \lc^2}{\kbt \lp}}\coth\left(\sqrt{\frac{f \lc^2}{\kbt \lp}}\right)-1\right]\, .
\ee
The equilibrium length for such a polymer is $\li=\lc^2/\lp$. Expanding $\ell(f)-\li$ to second order in the force $f$, and inverting the relation yields the force-extension relation to first anharmonic order for the semiflexible WLC:
\begin{eqnarray}
f_{\rm WLC}(x,\ve)=90 \kbt \left(\frac{\lp}{\lc^2}\right)^2(\ve \li-x)+\\
+\frac{5400}{7}\kbt \left(\frac{\lp}{\lc^2}\right)^3(\ve \li-x)^2\, .
\end{eqnarray}
So, with the identification
\begin{eqnarray}
\kt&=&90 \kbt \left(\frac{\lp}{\lc^2}\right)^2\\
\kd&=&\frac{5400}{7}\kbt \left(\frac{\lp}{\lc^2}\right)^3\, ,
\end{eqnarray}
we may use the external potential of Eq. (\ref{anharm}) to capture the lowest relevant nonlinear order of tethering by a semiflexible WLC. Obviously, for larger extensions (and thus forces) we may need to go to higher orders in $(\ve \li-x)$.

To verify the catch bonding effect predicted by Kramers for this anharmonic force extension relation on the mean unbinding time, We numerically solve Eq. (\ref{MFPT}) for a semiflexible polymer with $\ell_c=\ell_p=20$, confined to a trap with a spring constant strength of $\kt = 10$ and a dissociation length $b=1$. While we can no longer treat this case analytically, we do have access to the full range of $\ve$ allowing to compute, also, the total extent of the catch effect. Comparing the full WLC force-extension relation with its harmonic and first anharmonic approximations (potentials graphed in Fig. \ref{Fig3}(a)) confirms the Kramers predictions; to harmonic order; there is no catch bond effect but the inclusion of the first anharmonic term creates a regime of increasing lifetime with rising $\ve$. Thus, indeed, a nonlinear anharmonicity (third order terms or higher, odd powers, in $\Ue(x,\ve)$) are a prerequisite for this type of catch bond effect. The full WLC curve shows that the rise in lifetime continues over an extended range of $\ve$, and that the total induced lifetime increase can attain values of over $10\%$ (see Fig. \ref{Fig3}(b)).

So, both the Kramers analysis and the MFPT computations show a catch bonding effect for a particle in a symmetric, one-dimensional trap, attached to a polymer or, in fact, any tether possessing some anharmonic response. In the final part of this manuscript, we assess the real-life validity of the effect by a Molecular Dynamics (MD) simulation in three dimensions.

{\em Molecular Dynamics of the escape of a particle attached to a WLC in three dimensions}. We use LAMMPS MD \cite{Plimpton1995} to simulate a fluctuating semiflexible bead-spring chain consisting of $N$ beads, connected by identical Hookean springs with a spring constant $\kmd$ and rest length $\li$ and include a bending contribution to the chain energy, quantified by a bend stiffness $\sK$. The resulting chain energy is

\be
U_{\rm MD}=\sum_{i=1}^N\frac{\kmd}{2}(\ell_i-\ell_0)^2+\sum_{i=1}^{N-1}\frac{\sK}{2}\theta_i^2\, 
\ee
where $\ell_i$ the length is of spring $i$, and $\theta_i$ the angle between spring $i$ and spring $i+1$. One of the end beads is trapped in a harmonic potential with spherical symmetry, the other end is fixed at a distance $\ell=\li(1+\ve)$ from the center of the trap. In the simulation a polymer consisting of $N=40$ springs that each have a rest length $\ell_0=0.5$ (that is, the contour length of the entire polymer is $20$) and spring constant $\kmd=2500$ is used (all in Lennard-Jones units). This is very high, to supress backbone extension and approximate the inextensible WLC. The bending stiffness $\sK$ is set to $20$, corresponding to a persistence length $\ell_p$ of $20$ when we set $k_BT=1$.

\begin{figure}[t]
\begin{center}
\includegraphics[width=\linewidth]{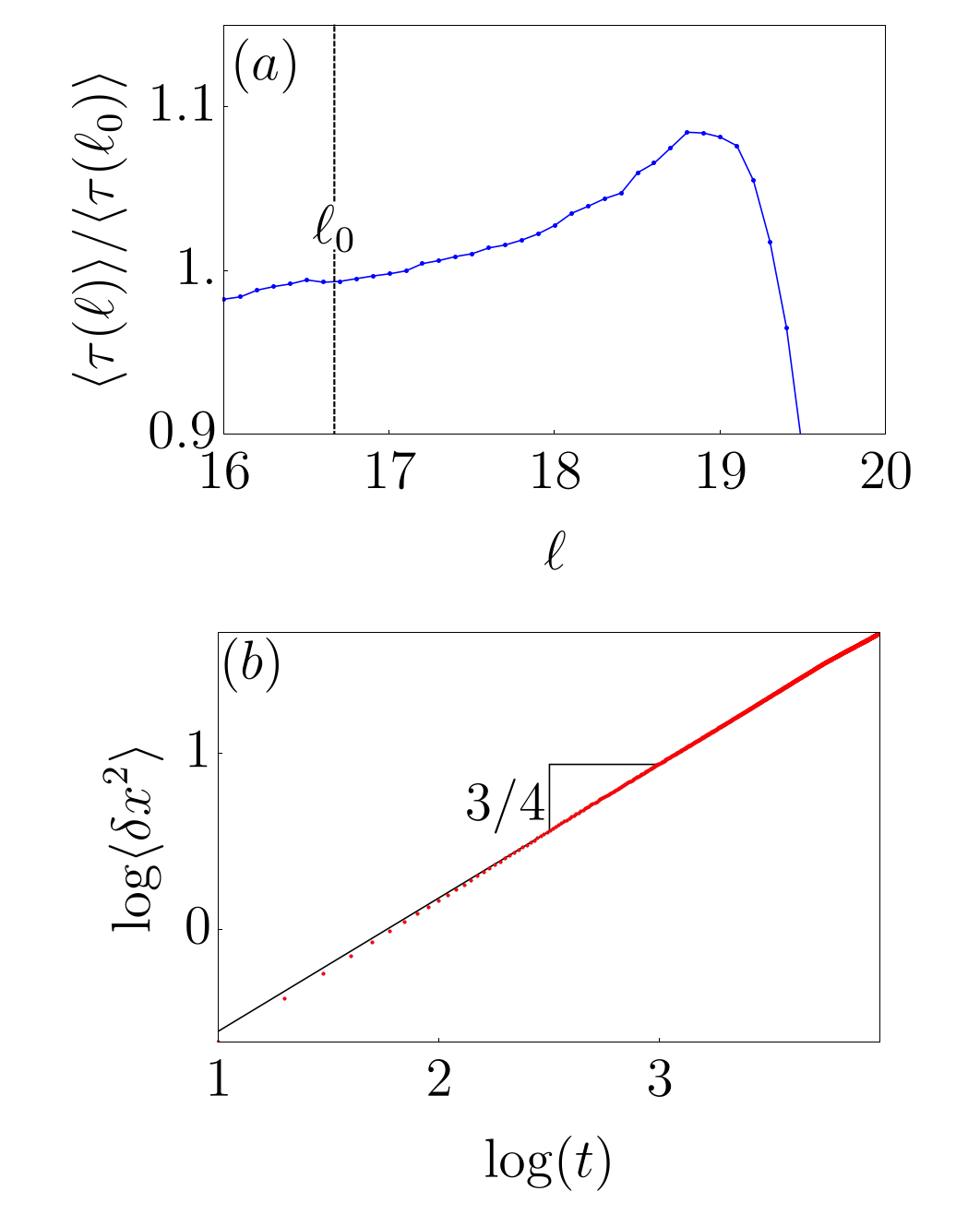}
\caption{\label{Fig4} (a) Normalized lifetime of particle attached to a bead-spring polymer obtained by MD simulation. (b) Mean square displacement of the end-bead showing subdiffusive motion.}
\end{center} 
\end{figure}

The three dimensional MD simulations again confirm the existence of the effect. As Fig. \ref{Fig4}(a) shows, the lifetime of the particle inside the trap rises with rising $\ve$, qualitatively in the manner seen in the MFPT apprach. There is, however, no satisfactory quantitative agreement between the numerical solution of the Eq. (\ref{MFPT}) and the MD simulations, despite the identical parameters. We believe this to be due to the anomalous diffusion of the polymer end point within the trap. As we show in Fig. \ref{Fig4}(b), the Mean Squared Displacement $\langle\delta x^2\rangle$ of the end bead scales as $\langle\delta x^2\rangle\sim t^\frac{3}{4}$, that is---slower than the expected power of 1. This subdiffusive motion of a polymer end point is well-known \cite{Hinczewski2009}, but invalidates the derivation that produces Eq. (\ref{MFPT}). To fully capture the escape problem of a polymer end point, we suggest a version of  Eq. (\ref{MFPT}) with fractional derivatives might be required, but this is beyond the scope of our current presentation.

{\em Summary and conclusion}
Our results prove, that catch bonding---bound states whose lifetime increases with applied force---is a generic feature of bonds where either the trap itself, or the external structures that a ligand is connected to, or both, display some degree of nonlinear response. These effects are robustly predicted by Kramers theory, MFPT theory, and are confirmed in direct MD simulation.

The mechanism we describe suggest catch bonding may be ubiquitous, and does not require finely tuned structural or conformational properties of bond participants. The same effects are predicted to occur in reversible (noncovalent) bonding in, for instance, polymeric materials; these reversible links should, under very general conditions, also show a regime of some strengthening when loaded. It will be instructive, in future analysis, to assess the theoretical limits of this effect to inspire the design of novel, responsive catch bond materials.    

\bibliography{ref}
\end{document}